\documentclass[12pt,preprint]{aastex6}

\usepackage{amsmath}
\usepackage{natbib}
\usepackage{xcolor}
\usepackage[]{graphics}
\usepackage{epsfig}
\usepackage{epstopdf}

\usepackage{mathrsfs}

\begin{document}

\title{Black Hole Shadow of Sgr A$^{*}$ in Dark Matter Halo}

\author{
   Xian Hou,\altaffilmark{1,2,3}
   Zhaoyi Xu,\altaffilmark{1,2,3,4}
   Ming Zhou,\altaffilmark{1,2,3}
  Jiancheng Wang \altaffilmark{1,2,3,4}
 }

\altaffiltext{1}{Yunnan Observatories, Chinese Academy of Sciences, 396 Yangfangwang, Guandu District, Kunming, 650216, P. R. China; {\tt xianhou.astro@gmail.com,zyxu88@ynao.ac.cn
}}
\altaffiltext{2}{Key Laboratory for the Structure and Evolution of Celestial Objects, Chinese Academy of Sciences, 396 Yangfangwang, Guandu District, Kunming, 650216, P. R. China}
\altaffiltext{3}{Center for Astronomical Mega-Science, Chinese Academy of Sciences, 20A Datun Road, Chaoyang District, Beijing, 100012, P. R. China}
\altaffiltext{4}{University of Chinese Academy of Sciences, Beijing, 100049, P. R. China}


\begin{abstract}
We study, for the first time, the shadow of the supermassive black hole Sgr A$^{*}$ at the center of the Milky Way in dark matter halos. For the Cold Dark Matter and Scalar Field Dark Matter models considered in this work, the apparent shape of the shadow depends upon the black hole spin $a$ and the dark matter parameter $k$. We find that both dark matter models influence the shadow in a similar way. The shadow is a perfect circle in the non-rotating case ($a=0$) and a deformed one in the rotating case ($a\neq0$). The size of the shadow increases with increasing $k$ in both non-rotating and rotating cases, while the shadow gets more and more distorted with increasing $a$ in the rotating case. We further investigate the black hole emission rate in both dark matter halos. We find that the emission rate decreases with increasing $k$ and the peak of the emission shifts to lower frequency. Finally, by calculating the angular radius of the shadow, we estimate that the dark matter halo could influence the shadow of Sgr A$^{*}$ at a level of order of magnitude of $10^{-3}$ $\mu$as and $10^{-5}$ $\mu$as, for CDM and SFDM, respectively. Future astronomical instruments with high angular resolution would be able to observe this effect and shed light on the nature of Sgr A$^{*}$. More interestingly, it may be possible to distinguish between CDM and SFDM models given the resolutions required differing by two orders of magnitude from each other. 

\end{abstract}


\section{INTRODUCTION}
It is widely believed that the center of our own galaxy, the Milky Way, hosts a supermassive black hole Sgr A$^{*}$. One of the ways to prove the existence of a black hole is to observe its shadow, which is the optical appearance cast by the black hole and appears as a two-dimensional dark zone for the observer on Earth. The black hole shadow can provide us information on fundamental properties of the black hole, i.e., the mass and the spin, thus enable a direct probe of the immediate environment of a black hole and the dynamics near the black hole. This will eventually serves as a test of fundamental predictions of Einstein's theory of General Relativity (GR). First images of the black hole Sgr A$^{*}$ at the center of the Milky Way and the black hole M87 at the center of the Virgo A galaxy are expected to be obtained using the sub-millimeter ``Event Horizon Telescope'' (EHT)\footnote{www.Eventhorizontelescope.org.} \citep{2008Natur.455...78D} based on the very-long baseline interferometry (VLBI).

The shadow of the Schwarzschild black hole was first discussed by \cite{1966MNRAS.131..463S} and later by \cite{1979A&A....75..228L} who considered the effect of a thin accretion disk on the shadow. \cite{1973blho.conf..215B} was the first to study the shadow cast by the Kerr black hole. Through constructing two observables, Hioki and Maeda \cite{2009PhRvD..80b4042H} examined the shadow of the Kerr black hole or a Kerr naked singularity. This topic has been extended to other black hole space-time by various researchers, e.g., Kerr-Newman black hole \citep{2000CQGra..17..123D,2005PASJ...57..273T,2018PhRvD..97f4021T}, Einstein-Maxwell-Dilation-Axion black hole\citep{2013JCAP...11..063W}, Kerr-Taub-NUT black hole \citep{2013Ap&SS.344..429A}, Braneworld black hole \citep{2009IJMPD..18..983S,2012PhRvD..85f4019A}, Kaluza-Klein rotating dilation black hole \citep{2013PhRvD..87d4057A}, non-Kerr black hole \citep[e.g.,][]{2012PhLB..711...10B,2013PhRvD..88f4004A,2017JCAP...10..051W}, Tomimatsu-Sato black hole \citep{2010CQGra..27t5006B}, Johannsen-Psaltis black hole \citep{2016PhRvD..94h4025Y}, Einstein-dilaton-Gauss-Bonnet black hole\citep{2016PhRvD..94h4025Y,2017PhLB..768..373C}, Kerr-Sen black hole \citep{2016arXiv161009477D,2016PhRvD..94h4025Y}, regular black holes \citep[e.g.,][]{2016PhRvD..93j4004A}, nonsingular black holes \citep[e.g.,][]{2016PhRvD..94b4054A}, Ay\'on Beato Garc\'ia black hole \citep{2018arXiv180203276S}, black hole with cosmological constant \citep{2014PhRvD..89l4004G,2018arXiv180404898P,2018EPJC...78...91E}, etc. Multiple shadows of a single black hole have also been discussed recently \citep[e.g.,][]{2015PhRvL.115u1102C,2018arXiv180203062G}, as well as the shadow of multiple black holes \citep{2012PhRvD..86j3001Y,2018arXiv180503798C}. The black hole shadow in modified GR has also been investigated in the literature, e.g., in extended Chern-Simons modified gravity \citep{2010PhRvD..81l4045A}, in Rastall gravity \citep{2017arXiv171209793K}, in Vector-Tensor Galileons Modified Gravity \citep{2018arXiv180104592V}, in fourth-order conformal Weyl gravity \citep{2017CaJPh..95.1299M}, etc. In addition, the study of shadow has been extended to black holes with higher or extra dimensions \citep[e.g.,][]{2014PhRvD..90b4073P, 2015EPJC...75..399A,2017arXiv170709521A,2017arXiv170707125P} and black holes surrounded by plasma \citep[e.g.,][]{2015PhRvD..92h4005A,2015PhRvD..92j4031P}. 
Specific attentions have been paid to the black hole Sgr A$^{*}$ using analytical approach and magnetohydrodynamic simulations considering more realistic situations with accretion flow and relativistic jets. The results have been compared with the EHT observations of Sgr A$^{*}$ to constrain the accretion and jet models  \citep[e.g.,][]{2000ApJ...528L..13F,2007CQGra..24S.259N,2010ApJ...717.1092D,2014A&A...570A...7M,2015ApJ...799....1C,2016ApJ...820..137B,2017ApJ...837..180G}. The possibility of testing theories of gravity basing on the shadow of Sgr A$^{*}$ has been equally discussed by various authors \citep[e.g.,][]{2006JPhCS..54..448B,2009PhRvD..79d3002B,2014ApJ...784....7B,2015ApJ...814..115P,2016PhRvL.116c1101J,2018NatAs.tmp...41M}. A review of the black hole shadow can be referred to \cite{2018GReGr..50...42C}.

On the other hand, according to the Standard Model of cosmology, the Universe is composed mostly of dark mater (27\%) and dark energy (68\%), while baryonic matter contributes only 5\% to the total mass-energy of the Universe. It is therefore natural to study the black hole shadow in the presence of dark energy and dark matter. Recently, \cite{2017arXiv171102898P} and \cite{2017IJMPD..2650051A} studied the black hole shadow in quintessence. Given that in galactic scale or near the black hole, the gravitational effect of dark matter is larger than dark energy, the influence of dark matter on the black hole properties might be equally more significant than dark energy. Thus, it is of greater interest to study the black hole shadow in dark matter halo.

Though no direct measurements have been made of the particle nature of dark matter, observational evidences supporting the existence of dark matter are accumulating through measurements of, for example, galactic rotation curves \citep{1980ApJ...238..471R}, galaxy cluster dynamics \citep{1933AcHPh...6..110Z}, the cosmic microwave background \citep{2014A&A...571A..16P}, the primordial abundances of heavy isotopes produced by big bang nucleosynethesis \citep{2003astro.ph..1505O}, etc. Among various theoretical dark matter models developed to explain the observations, Cold Dark Matter model (CDM) \citep{1996ApJ...462..563N,1997ApJ...490..493N,1991ApJ...378..496D} is the current most popular one which shows excellent consistence between observations and numerical simulations of large-scale structure of the Universe. However, strong tension exists between long-standing (and more recent) small-scale structure observations \citep{2018PhR...730....1T} and CDM model predictions.  Alternative models, such as Scalar Field Dark Matter model (SFDM) \citep[e.g.,][]{2000PhRvL..84.3760S,2002CQGra..19.6259U,2011JCAP...05..022H}, have been proposed. In particular, SFDM model can successfully, on the one hand, solve the small-scale structure problems and, on the other hand, keep great concordance with large-scale structure observations. 

A leading class of the dark matter particle candidates is Weakly Interacting Massive Particles (WIMPs), which are predicted to produce observable gamma rays, cosmic rays, and neutrinos through annihilations or decays. The Galactic center of the Milky Way, due to its proximity and high dark matter density, is expected to be the brightest dark matter source on the sky. Evidence of annihilation signal of WIMPs from the Galactic center has been arising in the past decade from gamma-ray observations with space and ground-based telescopes like {\it Fermi}-LAT and H.E.S.S \citep[See][for a review on this topic]{2015arXiv151102031C}. In this context, our work will, for the first time, investigate the shadow cast by the black hole Sgr A$^{*}$ at the Galactic center in dark matter halos.

The paper is organized as follows. In Section \ref{metric}, we introduce the space-time metric for spherical symmetric and rotating black hole in the two kinds of dark matter halos. In Section \ref{motion}, we derive the complete null geodesic equations and the motion of a test particle in the rotating black hole space-time. In Section \ref{shadow}, we study the apparent shapes of the shadow cast by the black hole Sgr A$^{*}$ in the presence of two kinds of dark matter halos. The energy emission rate of the black hole Sgr A$^{*}$ is investigated in Section \ref{emission} and we discuss our results in Section \ref{discussion}.

\section{BLACK HOLE SPACE-TIME IN DARK MATTER HALO}
\label{metric}
\subsection{Spherical symmetric black hole in dark matter halo}
The spherical symmetric black hole space-time metric in dark matter halo is \citep[][and references therein]{2018arXiv180300767X}
\begin{equation}
ds^{2}=-f(r)dt^{2}+\frac{dr^{2}}{f(r)}+r^{2}(d\theta^{2}+sin^{2}\theta d\phi^{2}).
\label{SBH1}
\end{equation}

For CDM halo:
\begin{equation}
f(r)=\left(1+\dfrac{r}{\tilde{R}}\right)^{-\dfrac{8\pi G\rho_{c}\tilde{R}^{3}}{c^{2}r}}- \dfrac{2GM}{rc^{2}},
\label{CDM1}
\end{equation}
where $M$ is the black hole mass, $c$ is the light of speed, $\rho_{c}$ is the density of the universe at the moment when the dark matter halo collapsed and $\tilde{R}$ is the characteristic radius. 

For SFDM halo:
\begin{equation}
f(r)=exp\left[-\dfrac{8G\rho_{c}R^{2}}{\pi}\dfrac{sin(\pi r/R)}{\pi r/R}\right]-\dfrac{2GM}{rc^{2}},
\label{SFDM1}
\end{equation}
where $\rho_{c}$ is the central density and $R$ is the radius at which the pressure and density are zero.

When there is no dark matter halo ($\rho_{c}=0$), the above metrics reduce to that of the Schwarzschild black hole.

\subsection{Rotating black hole in dark matter halo}
The rotating black hole space-time metric in dark matter halo is \citep{2018arXiv180300767X}
\begin{equation}
ds^{2}=-\left(1-\dfrac{r^{2}-f(r)r^{2}}{\Sigma^{2}}\right)dt^{2}+\dfrac{\Sigma^{2}}{\Delta}dr^{2}+\dfrac{2(r^{2}-f(r)r^{2})a sin^{2}\theta}{\Sigma^{2}}d\phi dt+$$$$
\Sigma^{2}d\theta^{2}+\dfrac{sin^{2}\theta}{\Sigma^{2}}((r^{2}+a^{2})^{2}-a^{2}\Delta sin^{2}\theta)d\phi^{2},
\label{KBH1}
\end{equation}
where
\begin{align}
&\Sigma^{2}=r^{2}+a^{2}cos^{2}\theta,\\
&\Delta=r^{2}f(r)+a^{2},
\label{KBH2}
\end{align}
and $f(r)$ takes the same form as Eq. (\ref{CDM1}) and Eq. (\ref{SFDM1}) for CDM and SFDM, respectively. Below we give the explicit expressions of the space-time metric for CDM halo and SFDM halo.

For CDM halo:
\begin{equation}
ds^{2}=-\left[1-\dfrac{r^{2}+\dfrac{2GMr}{c^{2}}-r^{2}\left(1+\dfrac{r}{\tilde{R}}\right)^{-\dfrac{8\pi G\rho_{c}\tilde{R}^{3}}{c^{2}r}}}{\Sigma^{2}}\right]dt^{2}+\dfrac{\Sigma^{2}}{\Delta}dr^{2}+
\Sigma^{2}d\theta^{2}+\dfrac{sin^{2}\theta}{\Sigma^{2}}((r^{2}+a^{2})^{2}-a^{2}\Delta sin^{2}\theta)d\phi^{2}$$$$
+\dfrac{2\left[r^{2}+\dfrac{2GMr}{c^{2}}-r^{2}\left(1+\dfrac{r}{\tilde{R}}\right)^{-\dfrac{8\pi G\rho_{c}\tilde{R}^{3}}{c^{2}r}}\right]a sin^{2}\theta}{\Sigma^{2}}d\phi dt,$$$$
\label{CDM2}
\end{equation}
where
\begin{equation}
\Delta=r^{2}\left(1+\dfrac{r}{\tilde{R}}\right)^{-\dfrac{8\pi G\rho_{c}\tilde{R}^{3}}{c^{2}r}}-\dfrac{2GMr}{c^{2}}+a^{2}.
\end{equation}

For SFDM halo:
\begin{equation}
ds^{2}=-\left[1-\dfrac{r^{2}+\dfrac{2GMr}{c^{2}}-r^{2}exp\left(-\dfrac{8G\rho_{c}R^{2}}{\pi}\dfrac{sin(\pi r/R)}{\pi r/R}\right)}{\Sigma^{2}}\right]dt^{2}+\dfrac{\Sigma^{2}}{\Delta}dr^{2}
+\Sigma^{2}d\theta^{2}$$$$
+\dfrac{2\left[r^{2}+\dfrac{2GMr}{c^{2}}
-r^{2}exp\left(-\dfrac{8G\rho_{c}R^{2}}{\pi}\dfrac{sin(\pi r/R)}{\pi r/R}\right)\right]a sin^{2}\theta}{\Sigma^{2}}d\phi dt
+\dfrac{sin^{2}\theta}{\Sigma^{2}}((r^{2}+a^{2})^{2}-a^{2}\Delta sin^{2}\theta)d\phi^{2},~~~$$$$
\label{SFDM2}
\end{equation}
where
\begin{equation}
\Delta=r^{2}exp\left[-\dfrac{8G\rho_{c}R^{2}}{\pi}\dfrac{sin(\pi r/R)}{\pi r/R}\right]-\dfrac{2GMr}{c^{2}}+a^{2}.
\end{equation}

In this work, we set $G=1$ and $c=1$. Furthermore, we define $k=\rho_{c}\tilde{R}^{3}$ and $k=\rho_{c}R^{3}$, for CDM halo and SFDM halo, respectively, to stand for the dark matter amount. For the black hole Sgr A$^{*}$, we adopt the values of different parameters reported in \cite{2015GReGr..47...12D}. For CDM halo, $\rho_{c}=1.936\times10^7$ M$_\odot$ kpc$^{-3}$ and $\tilde{R}=17.46$ kpc. For SFDM halo, $\rho_{c}=3.43\times10^7$ M$_\odot$ kpc$^{-3}$ and $R=15.7$ kpc. We then obtain $k=23965$ and $k=30869$ in unit of the mass of Sgr A$^{*}$ ($4.3\times10^6$ M$_\odot$), for CDM and SFDM, respectively. When the dark matter halo is absent ($\rho_{c}=0$ or $k=0$ equivalently), the above metrics reduce to that of the general Kerr black hole.

\section{NULL GEODESICS}
\label{motion}
Before studying the black hole shadow, it is necessary to first obtain the geodesic structure of a test particle for the above space-time metrics. For this, we employ the Hamilton-Jacobi equation and Carter constant separable method \citep{1968PhRv..174.1559C}. The Hamilton-Jacobi equation takes the general form as
\begin{equation}
\frac{\partial S}{\partial \sigma}=-\frac{1}{2}g^{\mu\nu}\frac{\partial S}{\partial {x^{\mu}}}\frac{\partial S}{\partial {x^{\nu}}}
\label{HJ1}
\end{equation}
where $S$ is the Jacobi action and $\sigma$ is an affine parameter along the geodesics. The separable solution of the Jacobi action $S$ reads as
\begin{equation}
S=\frac{1}{2}m^2\sigma-Et+L\phi+S_r(r)+S_{\theta}(\theta)
\label{HJ2}
\end{equation}
where $m$, $E$ and $L$ are, respectively, the test particle's mass, energy and angular momentum, with respect to the rotation axis. $S_r(r)$ and $S_{\theta}(\theta)$ are functions of $r$ and $\theta$, respectively. Inserting Eq. (\ref{HJ2}) into Eq. (\ref{HJ1}) and applying the variable separable method, we obtain the null geodesic equations for a test particle around the rotating black hole in dark matter halo as
\begin{align}
\label{HJ3}
&\Sigma\frac{dt}{d\sigma}=\frac{r^2+a^2}{\Delta}[E(r^2+a^2)-aL]-a(aE\sin^2\theta-L),\\
&\Sigma\frac{dr}{d\sigma}=\sqrt{\mathcal{R}},\\
&\Sigma\frac{d\theta}{d\sigma}=\sqrt{\Theta},\\
\label{HJ4}
&\Sigma\frac{d\phi}{d\sigma}=\frac{a}{\Delta}[E(r^2+a^2)-aL]-\left(aE-\frac{L}{\sin^2\theta}\right),
\end{align}
where $\mathcal{R}(r)$ and $\Theta(\theta)$ take the following form
\begin{align}
\label{HJ5}
&\mathcal{R}(r)=[E(r^2+a^2)-aL]^2-\Delta[m^2r^2+(aE-L)^2+\mathcal{K}],\\
&\Theta(\theta)=\mathcal{K}-\left(  \dfrac{L^2}{\sin^2\theta}-a^2E^2  \right) \cos^2\theta,
\end{align}
with $\mathcal{K}$ the Carter constant. The above geodesic equations (\ref{HJ3}-\ref{HJ4}) fully describe the dynamics of the test particle around the rotating black hole in dark matter halo. The boundary of the black hole shadow is mainly determined by the unstable circular orbit. We consider the case of photons and an observer at the infinity which implies that $m=0$ and photons arrive near the equatorial plane ($\theta=\pi/2$). The unstable circular orbit satisfies the condition
\begin{equation}
\mathcal{R}=\frac{\partial\mathcal{R}}{\partial r}=0.
\label{R1}
\end{equation}
Hence, from Eq. (\ref{HJ5}) and by introducing two impact parameters $\xi$ and $\eta$ 
\begin{equation}
\xi=L/E,  \;\; \;\; \;\;   \eta=\mathcal{K}/E^2,
\end{equation}
we obtain
\begin{align}
\label{R2}
&(r^2+a^2-a\xi)^2-[\eta+(\xi-a)^2](r^2f(r)+a^2)=0,\\
&4r(r^2+a^2-a\xi)-[\eta+(\xi-a)^2](2rf(r)+r^2f^{'}(r))=0.
\label{R3}
\end{align}
Combining Eqs. (\ref{R2}-\ref{R3}), we get the expressions of $\xi$ and $\eta$ as
\begin{align}
&\xi=\frac{(r^2+a^2)(rf'(r)+2f(r))-4(r^2f(r)+a^2)}{a(rf'(r)+2f(r))},\\
&\eta=\frac{r^3[8a^2f^{'}(r)-r(rf'(r)-2f(r))^2]}{a^2(rf'(r)+2f(r))^2}.
\end{align}
Furthermore, we have
\begin{align}
\xi^2+\eta &=2r^2+a^2+\frac{16(r^2f(r)+a^2)}{(rf'(r)+2f(r))^2}-\frac{8(r^2f(r)+a^2)}{rf'(r)+2f(r)}\\
                &=2r^2+a^2+\frac{8\Delta[2-(rf'(r)+2f(r)]}{(rf'(r)+2f(r))^2}.
\end{align}

For the CDM halo (Eq. \ref{CDM1}), we have
\begin{equation}
f'(r)=\left(1+\frac{r}{\tilde{R}}\right)^{-\dfrac{8\pi k}{r}}\left[\frac{8\pi k}{r^2}\ln\left(1+\frac{r}{\tilde{R}}\right)- \frac{8\pi k}{r(r+\tilde{R})} \right]+\frac{2M}{r^2},
\end{equation}
and then
\begin{equation}
\xi^2+\eta = 2r^2+a^2 +  \dfrac{8\left[ r^2\left( 1+\dfrac{r}{\tilde{R}} \right)^{-\dfrac{8\pi k}{r}}  -2Mr+a^2   \right]  \left[ 2+\dfrac{2M}{r}- \left(1+\dfrac{r}{\tilde{R}} \right)^{-\dfrac{8\pi k}{r}} \left( \dfrac{8\pi k}{r} \ln\left(1+\dfrac{r}{\tilde{R}}\right)- \dfrac{8\pi k}{(r+\tilde{R})}   +2   \right)  \right]  }      { \left[  \left(1+\dfrac{r}{\tilde{R}} \right)^{-\dfrac{8\pi k}{r}} \left( \dfrac{8\pi k}{r} \ln\left(1+\dfrac{r}{\tilde{R}}\right)- \dfrac{8\pi k}{(r+\tilde{R})}   +2   \right)   -   \dfrac{2M}{r}         \right]^2               }
\end{equation}

Similarly, for the SFDM halo (Eq. \ref{SFDM1}), we have 
\begin{equation}
f'(r)=-\frac{8k}{\pi^2}\left[\frac{\pi}{Rr}\cos\left(\frac{\pi r}{R}\right)-\frac{1}{r^2}\sin\left(\dfrac{\pi r}{R}      \right)  \right] exp\left[-\dfrac{8k}{\pi}\dfrac{sin(\pi r/R)}{\pi r}\right] +\frac{2M}{r^2},
\end{equation}
and then
\begin{equation}
\xi^2+\eta = 2r^2+a^2 $$$$
+\dfrac{   8\left[  r^2 exp\left(-\dfrac{8k}{\pi}\dfrac{sin\left(\pi r/R\right)}{\pi r} \right)  -2Mr+a^2 \right]   \left[  2+\dfrac{2M}{r}-  \left( \dfrac{8k}{\pi^2r}\sin \left(\dfrac{\pi r}{R}\right) -\dfrac{8k}{\pi R}  \cos\left(\dfrac{\pi r}{R}\right)  +2 \right) exp\left(-\dfrac{8k}{\pi}\dfrac{sin(\pi r/R)}{\pi r}  \right)   \right]   }        {\left[ \left( \dfrac{8k}{\pi^2r}\sin \left(\dfrac{\pi r}{R}\right) -\dfrac{8k}{\pi R}  \cos\left(\dfrac{\pi r}{R}\right) +2 \right) exp\left(-\dfrac{8k}{\pi}\dfrac{sin(\pi r/R)}{\pi r}  \right)  -\dfrac{2M}{r} \right]^2}
\end{equation}

\section{BLACK HOLE SHADOW}
\label{shadow}
To determine the shape of the black hole shadow, we introduce the celestial coordinates $\alpha$ and $\beta$ as 
\begin{align}
& \alpha = \lim_{r_o\to \infty}\left( -r_o^2 \sin \theta_o \dfrac{d\phi}{dr}  \right),\\
& \beta = \lim_{r_o \to \infty}\left( r_o^2 \dfrac{d\theta}{dr}  \right),
\end{align}
where $r_o$ is the distance between the black hole and the observer, $\theta_o$ is the angle between the rotation axis of the black hole and the line of sight of the observer (i.e., inclination angle). Here we assume the observer is at infinity. $\alpha$ is the apparent perpendicular distance of the shadow as seen from the axis of symmetry, and $\beta$ is the apparent perpendicular distance of the shadow as seen from its projection on the equatorial plane.

Using the null geodesic equations (\ref{HJ3}-\ref{HJ4}), we can obtain the relations between celestial coordinates and impact parameters $\xi$ and $\eta$ as
\begin{align}
& \alpha = -\dfrac{\xi}{sin \theta},\\
& \beta = \pm \sqrt{\eta + a^2\cos^2\theta -\xi^2\cot^2 \theta}.
\end{align}

In the equatorial plane ($\theta=\pi/2$), $\alpha$ and $\beta$ reduce to
\begin{align}
& \alpha = -\xi,\\
& \beta = \pm \sqrt{\eta }.
\end{align}

\begin{figure}[htbp]
  \centering
   \includegraphics[scale=0.5]{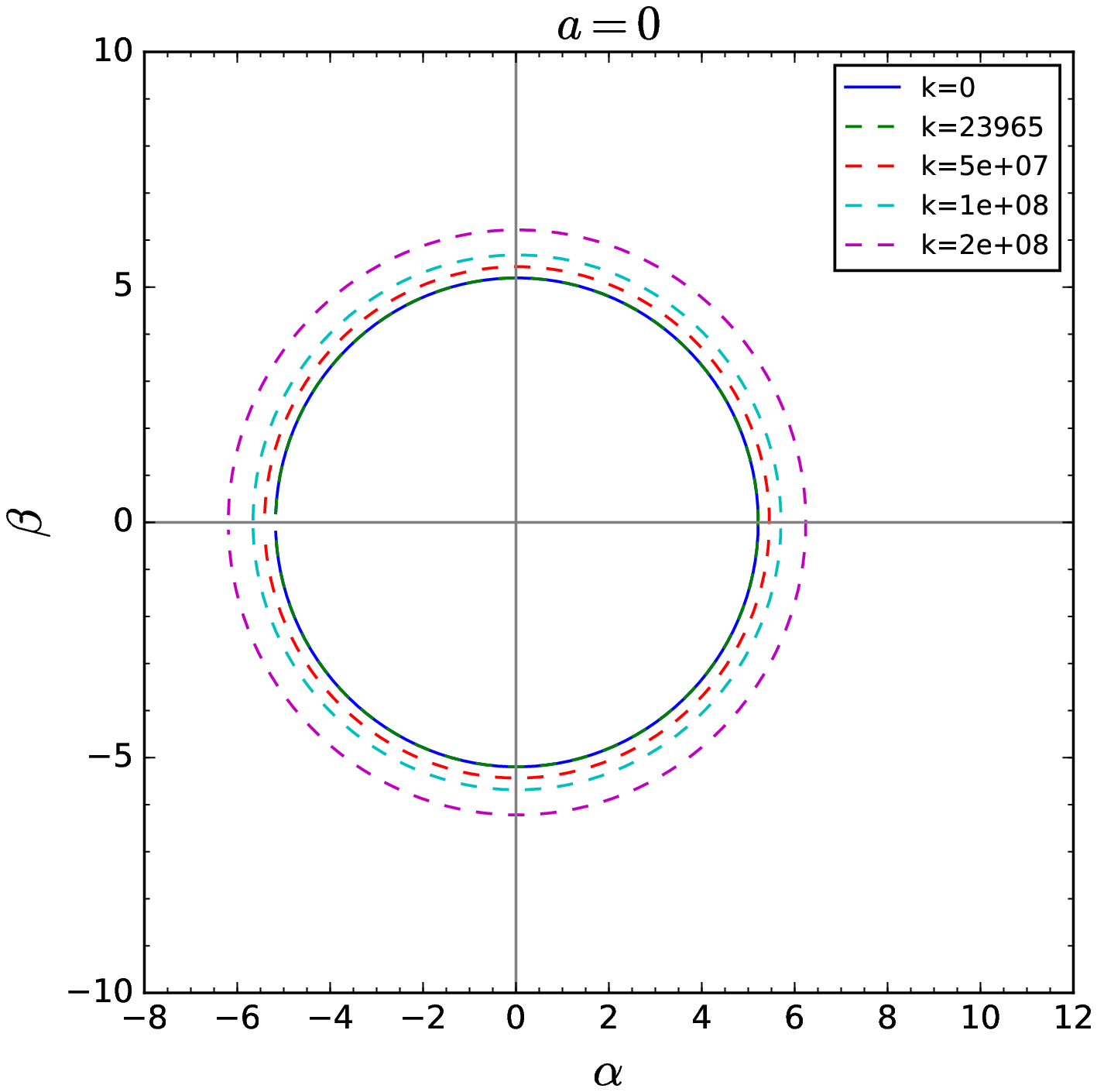}
   \includegraphics[scale=0.5]{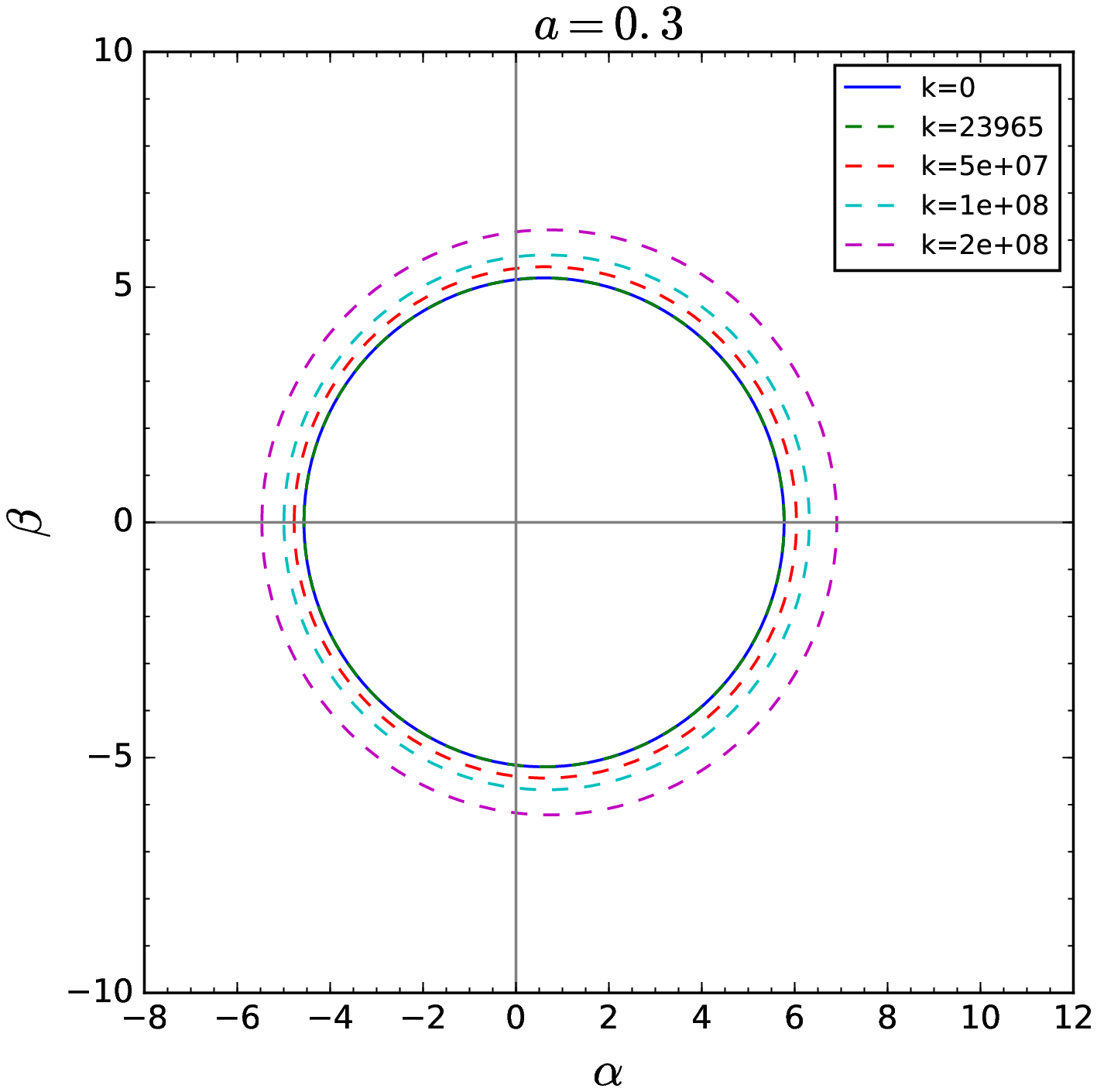}
   \includegraphics[scale=0.5]{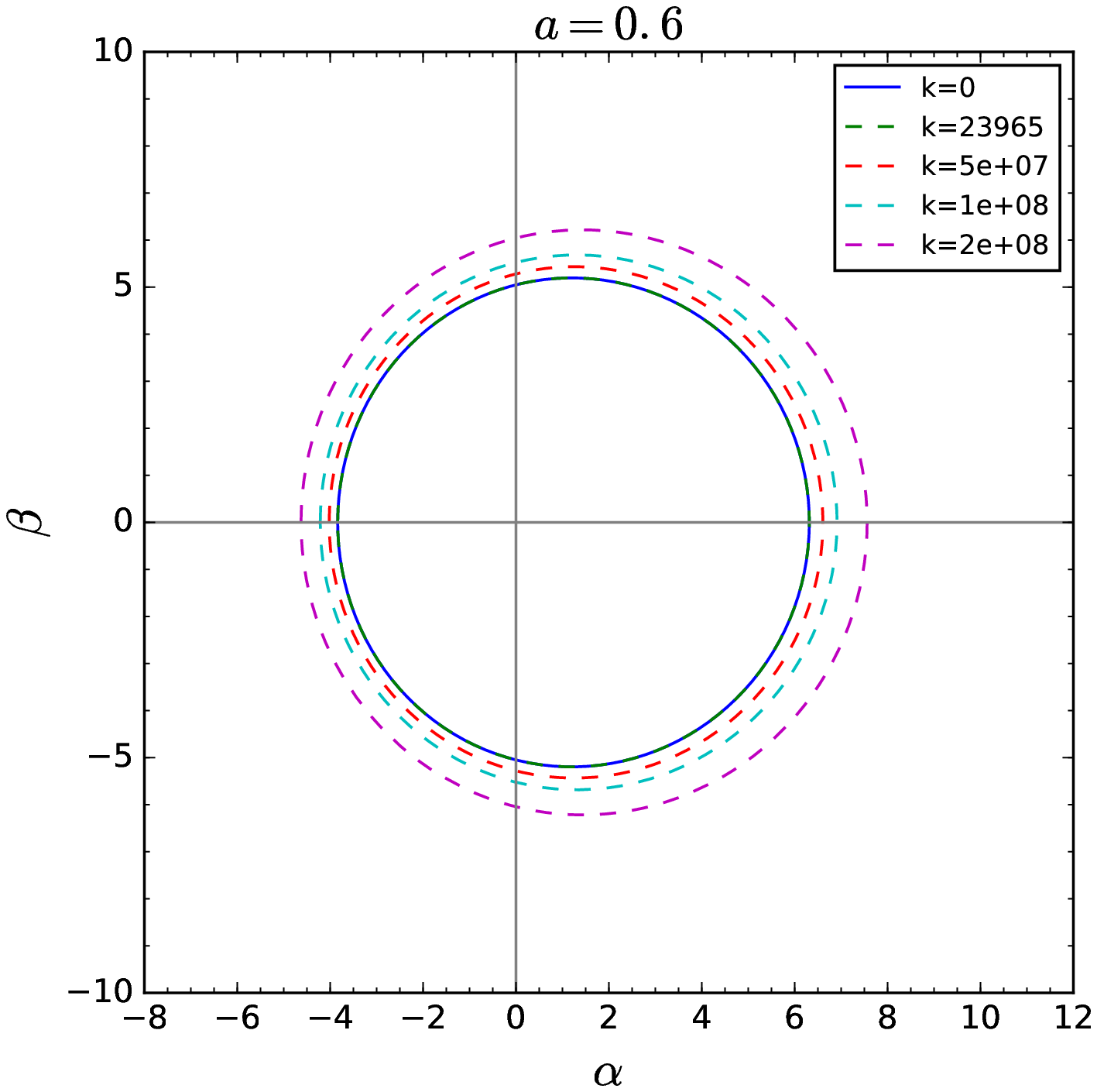}
   \includegraphics[scale=0.5]{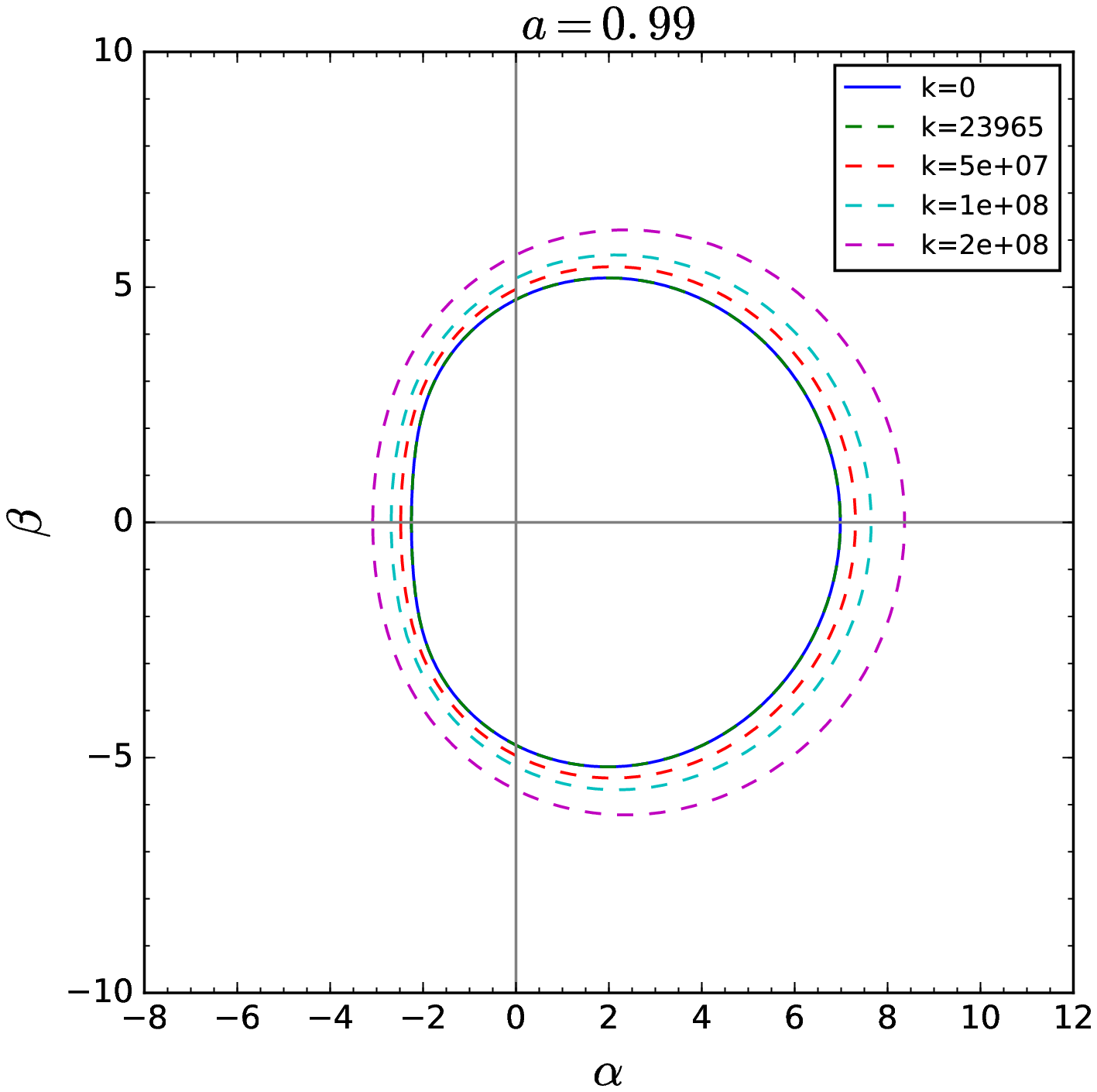}
   \caption{Silhouette of the shadow cast by the rotating black hole Sgr A$^{*}$ in the CDM halo for different values of parameters $a$ and $k$.}
  \label{shadow_CDM}
\end{figure}

\begin{figure}[htbp]
  \centering
   \includegraphics[scale=0.5]{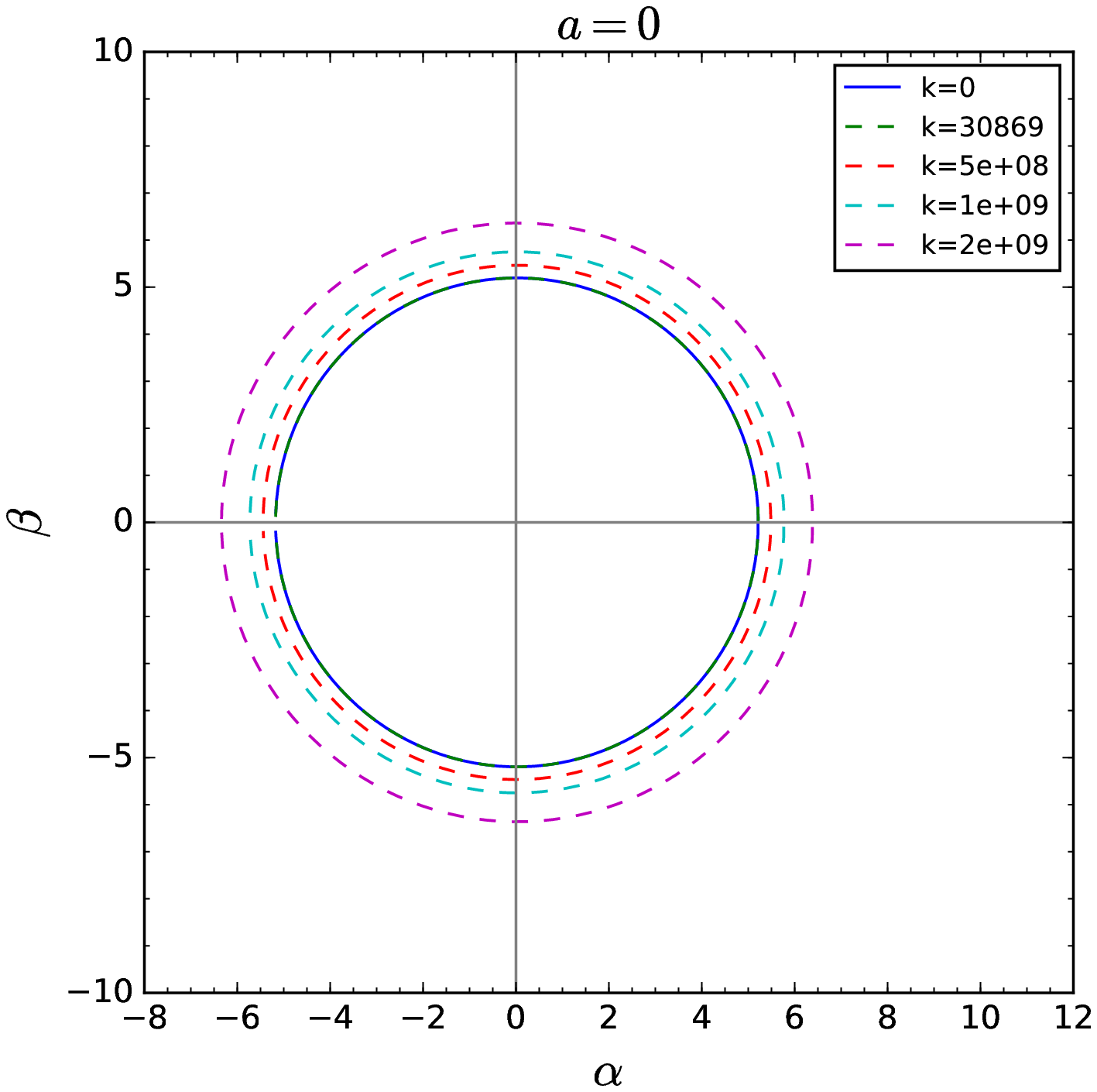}
   \includegraphics[scale=0.5]{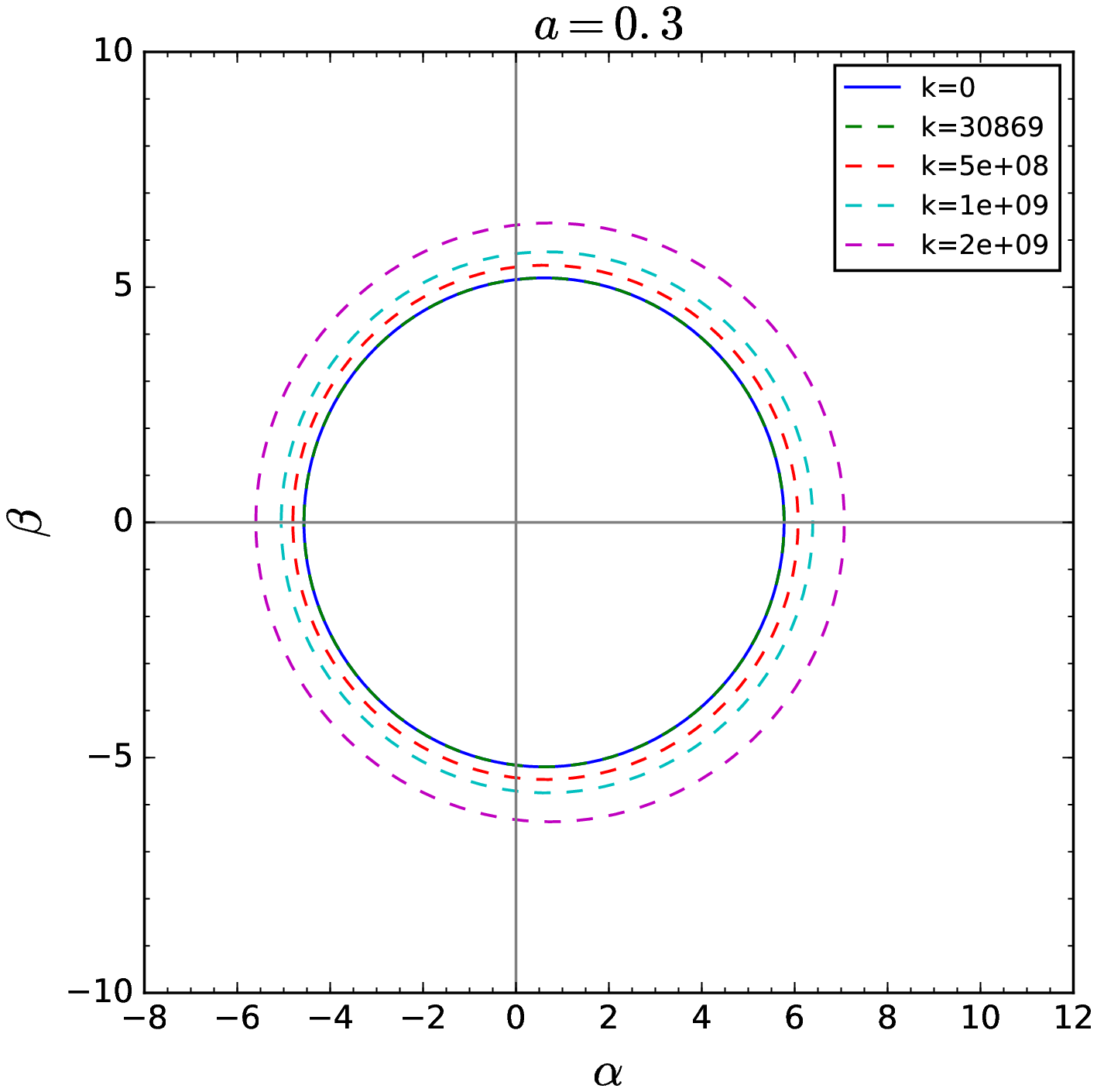}
   \includegraphics[scale=0.5]{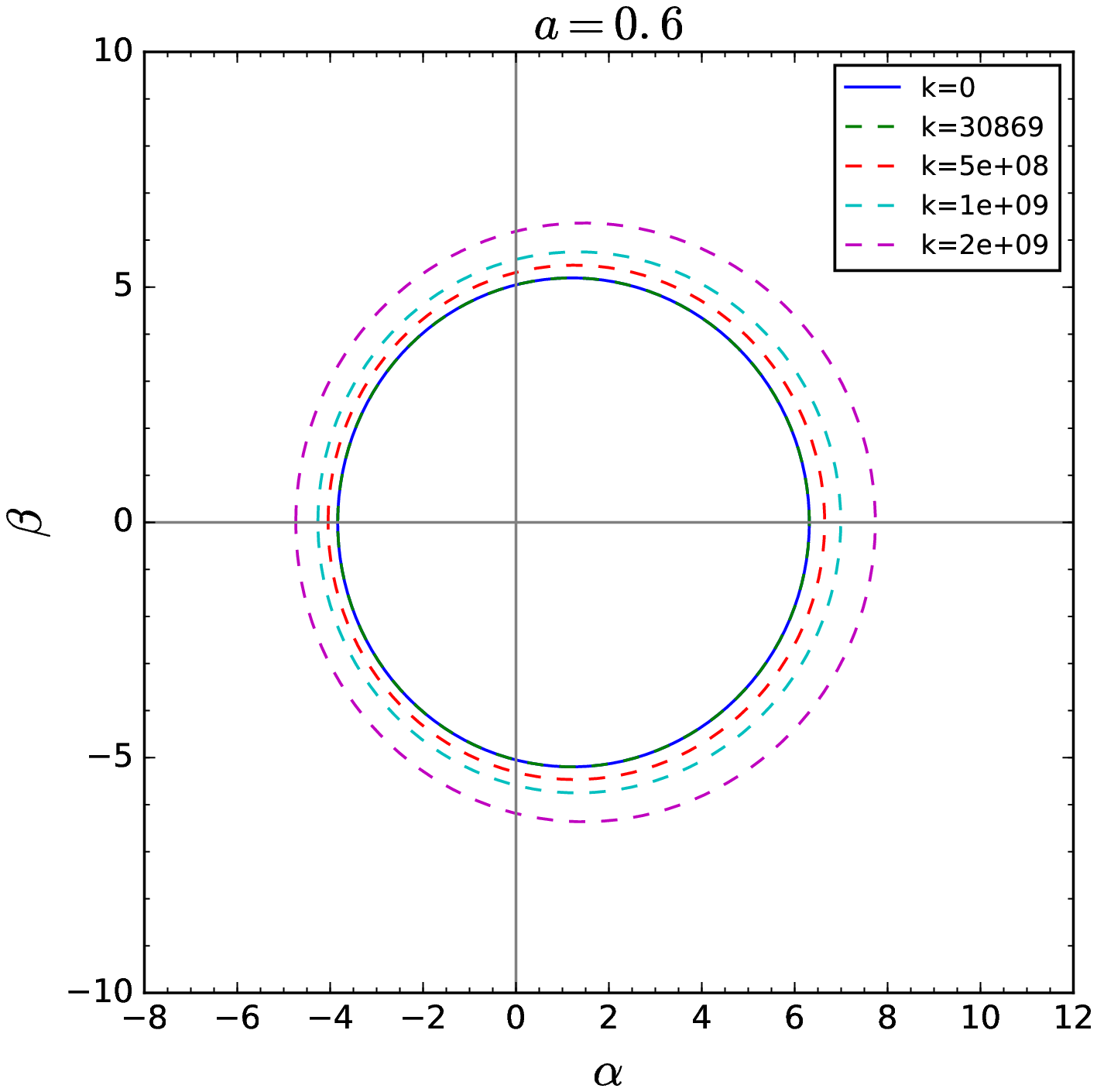}
   \includegraphics[scale=0.5]{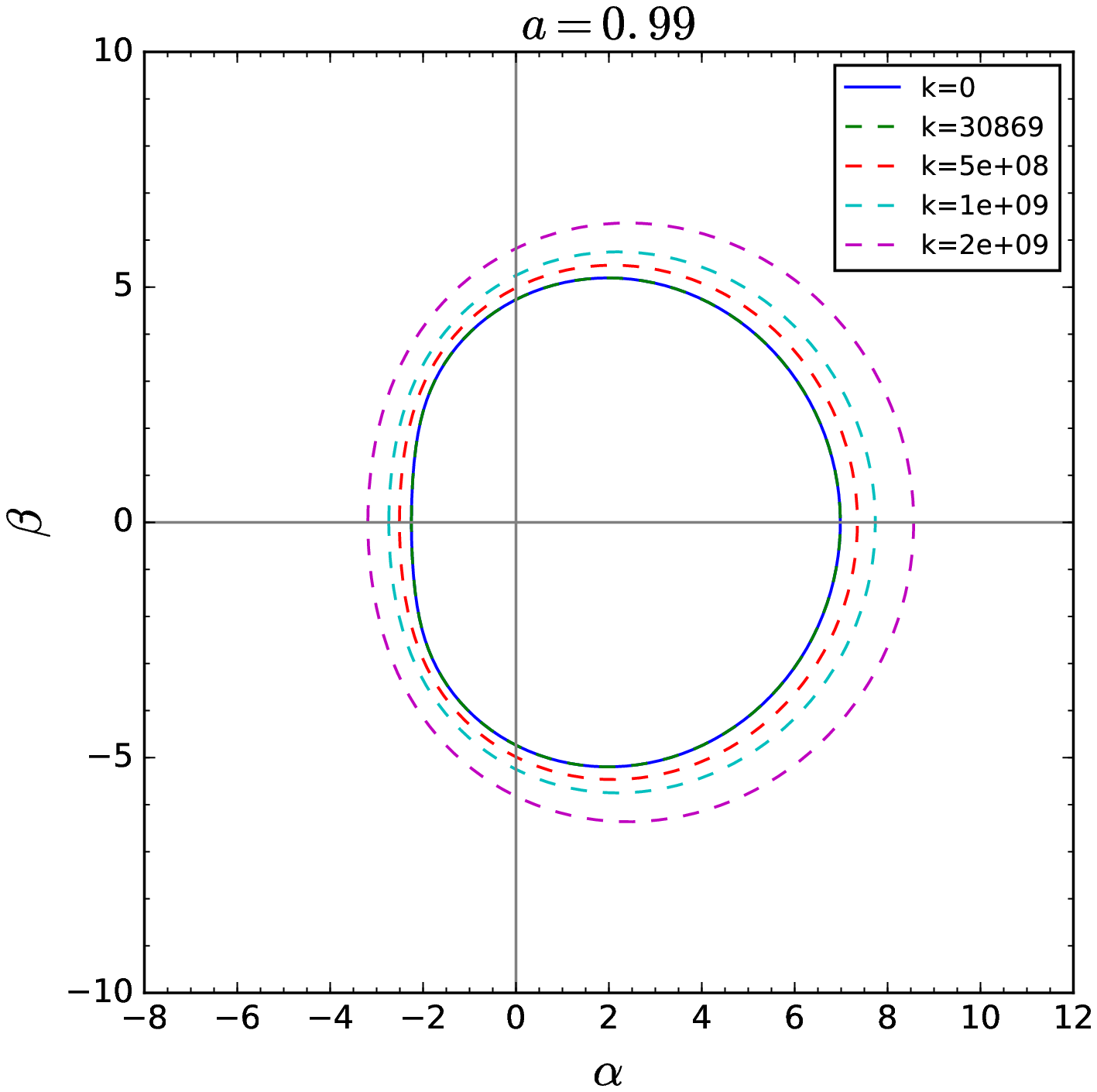}
   \caption{Silhouette of the shadow cast by the rotating black hole Sgr A$^{*}$ in the SFDM halo for different values of parameters $a$ and $k$.}
  \label{shadow_SFDM}
\end{figure}

By plotting $\beta$ against $\alpha$, we show different shapes of the shadow in Figure \ref{shadow_CDM} and \ref{shadow_SFDM}, for the CDM halo and SFDM halo, respectively. In the non-rotating case ($a=0$), the shadow is a perfect circle and the size increases with increasing $k$. In the rotating case ($a\neq 0$), the shadow gets more and more distorted for a higher $a$ given fixed $k$, and the size increases with increasing $k$ given fixed $a$, similar to the case of $a=0$. We note that the influence of dark matter on the shadow is actually minor with a visible effect only when $k$ increases to order of magnitude of $10^7$.

From our study, the structure of the black hole shadow in the CDM and SFDM halos is very similar to the cases of Schwarzschild and Kerr black hole. We propose that this is because in our case, the fundamental properties of the black hole is maintained, while the dark matter halo only induces a fluctuation-like effect on the shadow. We can find similar phenomena for black hole shadows in plasma. Yet, if the black hole under study is not Schwarzschild or Kerr black hole, but with extra fundamental parameters, like dilation, or in alternative theories of gravity, the shadow would rather exhibit some obvious dissimilarities. An example of comparison of different black hole shadows can be found in \cite{2017IJMPD..2630001G}, Figure 7.

\begin{figure}[htbp]
  \centering
  \includegraphics[scale=0.5]{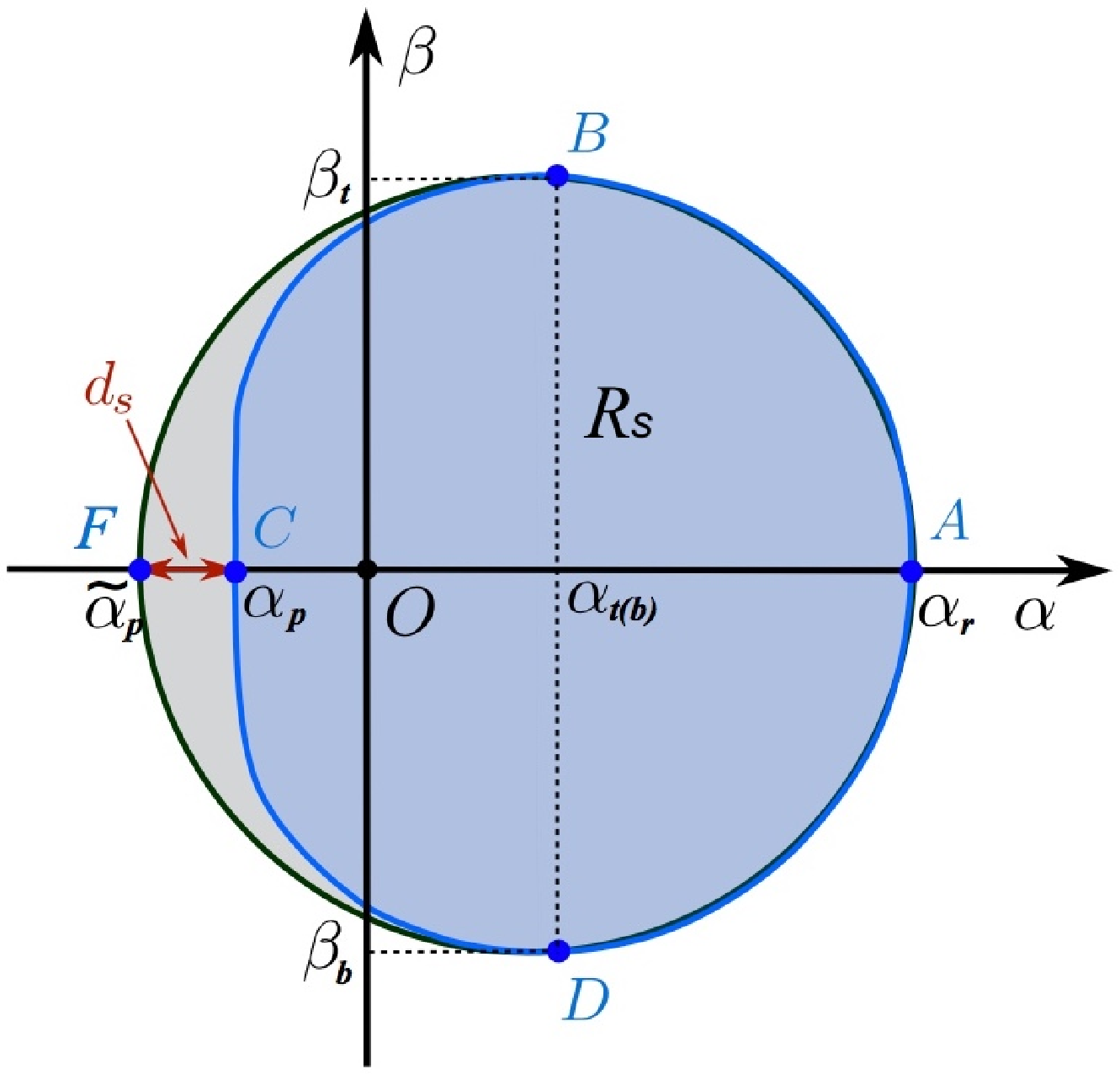}
   \caption{Schematic illustration of the black hole shadow and the reference circle.}
  \label{shadow_ref}
\end{figure}

Further study of the black hole shadow relies on two astronomical observables defined in \cite{2009PhRvD..80b4042H}: the radius of the shadow $R_s$ and the distortion parameter $\delta_s$. The schematic illustration of $R_s$ and $\delta_s$ is shown in Figure \ref{shadow_ref}. $R_s$ is the radius of the reference circle passing through three points: the top one $B(\alpha_t, \beta_t)$, the bottom one  $D(\alpha_b, \beta_b)$ and the most right one $A(\alpha_r, 0)$. The points $C(\alpha_p, 0)$ and $F(\tilde{\alpha}_p, 0)$ are where the circle of the shadow and the reference circle cut the horizontal axis at the opposite side of $A(\alpha_r, 0)$, respectively. $d_s$ is the distance from the most left position (C) of the shadow to the reference circle (F). $R_s$ approximately gives the size of the shadow and $\delta_s$ measures its deformation with respect to the reference circle. From the geometry of the shadow, we have
\begin{equation}
R_s = \dfrac{(\alpha_t-\alpha_r)^2 + \beta_t^2}{2|\alpha_r-\alpha_t|},
\end{equation}
where we have used the relations $\alpha_b=\alpha_t$ and $\beta_b=-\beta_t$. And
\begin{equation}
\delta_s = \dfrac{d_s}{R_s} = \dfrac{|\alpha_p-\tilde{\alpha}_p|}{R_s}.
\end{equation}

Considering the relation $\tilde{\alpha}_p=\alpha_r-2 R_s$, we have
\begin{equation}
\delta_s = 2-\dfrac{D_s}{R_s}
\end{equation}
where $D_s=\alpha_r-\alpha_p$ is the diameter of the shadow along the axis of $\alpha$.

In the non-rotating case ($a=0$), the shadow of the black hole is a perfect circle with radius of $R_s$. So
\begin{equation}
\alpha^2 + \beta^2 = \xi^2 +\eta= R_s^2. 
\end{equation}

\begin{figure}[htbp]
  \centering
   \includegraphics[scale=0.45]{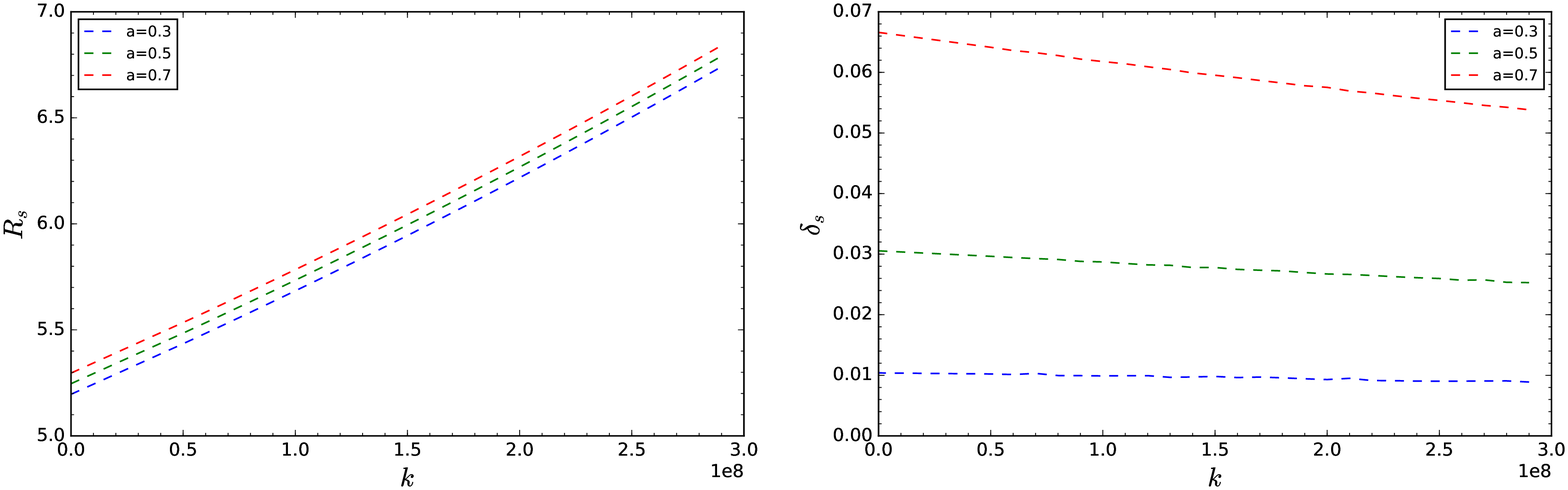}
   \caption{Variation of the radius $R_s$ (left) and the distortion parameter $\delta_s$ (right) of the shadow of Sgr A$^{*}$ with the parameters $a$ and $k$ in the CDM halo. The lines of $R_s$ have been moved up vertically to visualize the trend of $R_s$ for different $a$ by adding a constant to $R_s$.}
  \label{Rs_CDM}
\end{figure}

\begin{figure}[htbp]
  \centering
   \includegraphics[scale=0.45]{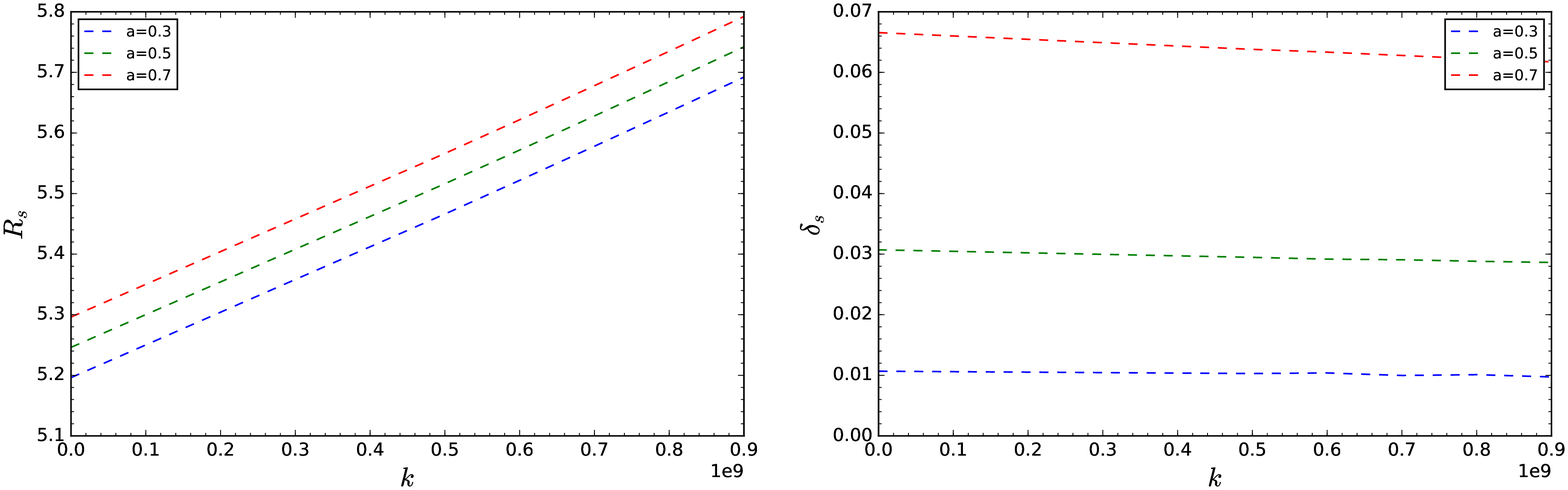}
   \caption{Variation of the radius $R_s$ (left) and the distortion parameter $\delta_s$ (right) of the shadow of Sgr A$^{*}$ with the parameters $a$ and $k$ in the SFDM halo. The lines of $R_s$ have been moved up vertically to visualize the trend of $R_s$ for different $a$ by adding a constant to $R_s$.}
  \label{Rs_SFDM}
\end{figure}

Figure \ref{Rs_CDM} and \ref{Rs_SFDM} show the variation of the radius $R_s$ and the distortion parameter $\delta_s$ with the parameters $a$ and $k$ for the CDM halo and SFDM halo, respectively. We find that the radius of the shadow increases with the increasing $k$, but almost does not vary with $a$ (a constant has been added to visualize the trend of $R_s$ for different $a$). The distortion parameter decreases monotonically with increasing $k$ for a given $a$, and increases with $a$ for a given $k$. This trend could also be inferred from Figure \ref{shadow_CDM} and \ref{shadow_SFDM}. Similarly, the effect of dark matter is visible only when $k$ increases to order of magnitude of $10^7$.

\section{ENERGY EMISSION RATE}
\label{emission}
For an observer located at an infinite distance, the black hole shadow corresponds to the high energy absorption cross section, the latter oscillating around a limiting constant value $\sigma_{lim}$ for a spherical symmetric black hole. $\sigma_{lim}$ is approximately equal to the geometrical cross section of the photon sphere \citep{1973PhRvD...7.2807M,1973grav.book.....M} and can be expressed as \citep{2013JCAP...11..063W} 
\begin{equation}
\sigma_{lim} \approx \pi R_s^2,
\end{equation}
with $R_s$ the radius of the black hole shadow. This can be generalized to the rotating black hole considered in this work, given that the shadow approaches to a standard circle as can be seen from Figure \ref{shadow_CDM} and \ref{shadow_SFDM}. The energy emission rate of the black hole is therefore 
\begin{equation}
\dfrac{d^2E(\omega)}{d\omega dt} = \dfrac{2\pi^2\sigma_{lim} }{e^{\omega/T}-1}\omega^3
\end{equation}
with $\omega$ the frequency of photon and $T$ the Hawking temperature for the outer event horizon which is defined by
\begin{equation}
T = \lim_{\theta=0,  r \to r_+} \dfrac{\partial_r \sqrt{g_{tt}}   }{2\pi \sqrt{g_{rr}}}.
\end{equation}
In our case, we have for the rotating black hole in dark matter halo
\begin{equation}
g_{tt} = 1-\dfrac{r^{2}-f(r)r^{2}}{\Sigma^{2}},  ~~~~~ g_{rr} = \dfrac{\Sigma^{2}}{\Delta}.
\end{equation}

Thus, we obtain the Hawking temperature as
\begin{equation}
T = \dfrac{r_+^2f'(r_+)(r_+^2+a^2) + 2a^2r_+(f(r_+)-1)  }{ 4\pi (r_+^2+a^2)^2      }
\label{T_hawk}
\end{equation} 
where $r_+$ is the outer event horizon of the black hole defined as the greater root of the solution for $1/g_{rr}=0$.

Eq. (\ref{T_hawk}) reduces to the regular Kerr black hole in the case of $k=0$ and takes the form
\begin{equation}
T_{Kerr} = \dfrac{r_+^2-a^2 }{ 4\pi r_+ (r_+^2+a^2) }
\end{equation} 
with $r_+ = M+\sqrt{M^2-a^2}$.

\begin{figure}[htbp]
  \centering
   \includegraphics[scale=0.5]{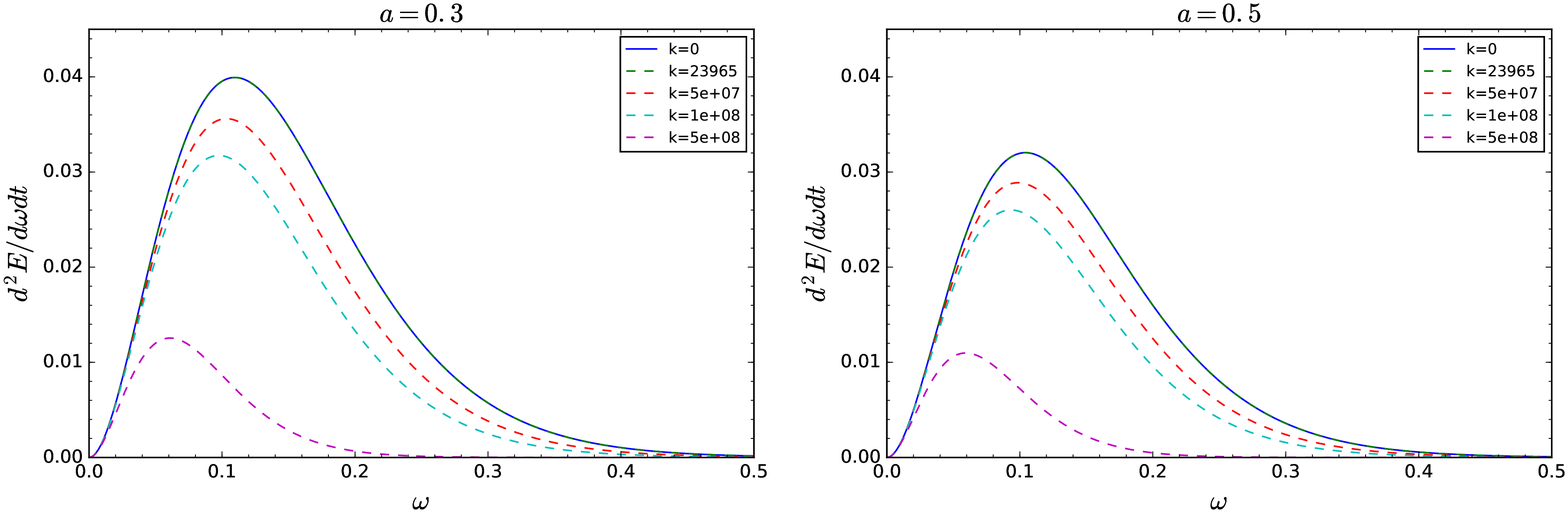}
   \caption{Evolution of the emission rate with the frequency $\omega$ for different values of the parameters $a$ and  $k$ for the CDM halo.}
  \label{Erate_CDM}
\end{figure}

\begin{figure}[htbp]
  \centering
   \includegraphics[scale=0.5]{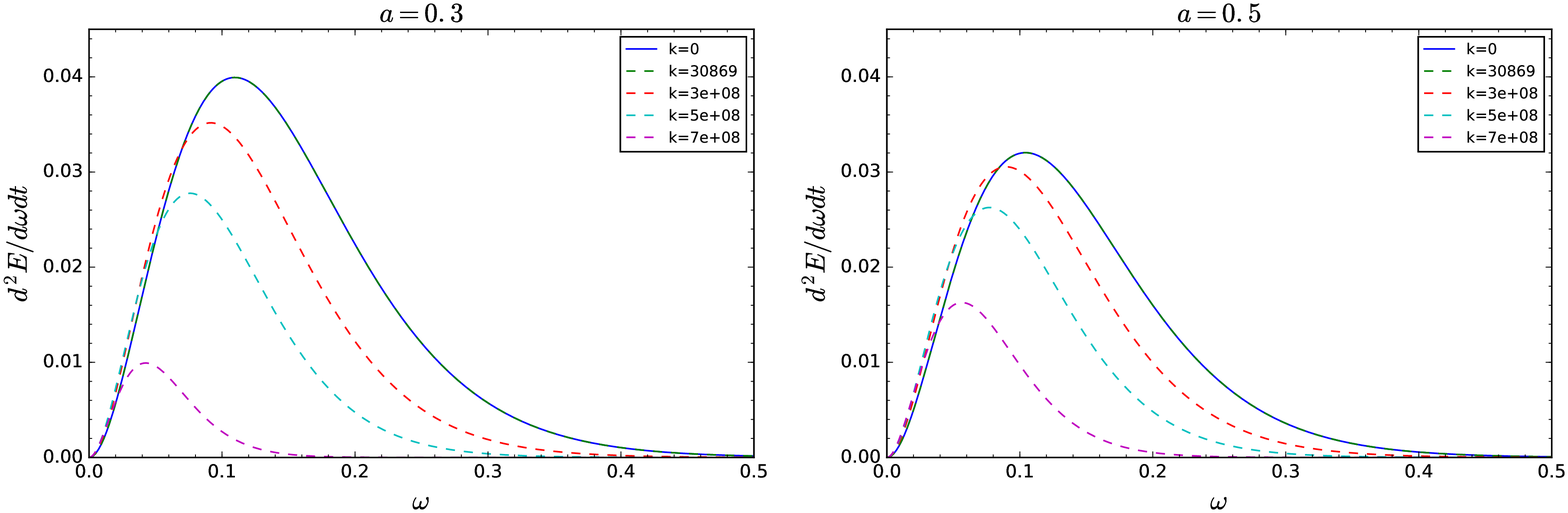}
   \caption{Evolution of the emission rate with the frequency $\omega$ for different values of the parameters $a$ and $k$ for the SFDM halo.}
  \label{Erate_SFDM}
\end{figure}

In Figure \ref{Erate_CDM} and  \ref{Erate_SFDM}, we show the energy emission rate against the frequency $\omega$ for different values of the parameters $a$ and $k$, assuming a CDM halo and SFDM halo, respectively. We can see that the peak of the emission decreases with increasing $k$ and shifts to lower frequency. Similarly, the effect of dark matter is minor and only visible when $k$ increases to order of magnitude of $10^7$.

\section{DISCUSSION}
\label{discussion}
In this work, we study the shadow cast by the black hole Sgr A$^{*}$ at the center of the Milky Way in dark matter halo by analysing how the shadow is influenced by the black hole spin $a$ and the dark matter parameter $k$. We find that the two dark matter models (CDM and SFDM) considered in this work affect the shadow in a similar way for an observer located at infinity and in the equatorial plane. For a fixed value of $a$, the size of the shadow $R_s$ increases with increasing $k$ and the distortion parameter $\delta_s$ monotonically decreases. For a fixed value of $k$, the shadow gets more and more distorted with increasing $a$, characterized by larger and larger $\delta_s$. With the assumption that the black hole shadow equals to the high energy absorption cross section, we calculate the emission rate of Sgr A$^{*}$ in dark matter halo. We find that for both dark matter models, the emission rate decreases with increasing $k$ for a fixed frequency $\omega$ and the peak of the emission shifts to lower $\omega$. In general, the influence of dark matter on the black hole is minor and only becomes significant when $k$ increases to order of magnitude of $10^7$, for both CDM and SFDM models.

The angular radius of the shadow can be estimated using the observable $R_s$ as $\theta_s = R_sM/D$, where $M$ is the black hole mass and $D$ is the distance between the black hole and the observer. The angular radius can be further expressed as $\theta_s = 9.87098\times 10^{-6} R_s(M/M\odot)(1 \rm{kpc}/$ $D)$ $\,\mu$as \citep{2012PhRvD..85f4019A}. For the supermassive black hole Sgr A$^{*}$ at the center of the Milky Way, its mass is estimated to be $M=4.3\times10^6 M\odot$ and $D=8.3$ kpc which is the distance between the Earth and the black hole. Through calculations for $R_s$ and $\theta_s$, the angular resolution required to detect the dark matter influence on the black hole shadow would be, for CDM, $10^{-3}$ $\mu$as (i.e., to distinguish between $k=0$ and $k=23965$) and for SFDM, $10^{-5}$ $\mu$as (i.e., to distinguish between $k=0$ and $k=30869$). This is out of the reach of the current astronomical instruments. For example, the current EHT resolution is $\sim$ 60 $\mu$as at 230 GHz and will be able to achieve a finer one of 15 $\mu$as by observing at a higher frequency of 345 GHz and adding more VLBI telescopes. The space-based VLBI RadioAstron \citep{2013ARep...57..153K}\footnote{http://www.asc.rssi.ru/radioastron/index.html} will be able to obtain a resolution of $\sim 1-10$ $\mu$as.  This is still at least three orders of magnitude lower than the resolution required by the CDM model. The angular resolution of a baseline is given by  $1 / f D$, with $f$ the observing frequency and $D$ the baseline which is the separation of EHT sites used for VLBI. We estimated that, for example, to achieve the resolution of $10^{-3}$ $\mu$as required by the CDM model, observations at a much higher frequency of $\sim5\times10^6$ GHz or with a much longer baseline of $\sim 4\times10^8$ km will be needed.

We anticipate that future observations with highly improved techniques would be able to achieve the resolution required to observe the dark matter influence on the shadow of the black hole Sgr A$^{*}$. Furthermore, the angular resolution difference between CDM and SFDM models is as large as two orders of magnitude. This implies that observing the black hole shadow of Sgr A$^{*}$ may serve as a tool of distinguishing one model from the other and eventually shed light on the nature of Sgr A$^{*}$ and dark matter.


\acknowledgments
We acknowledge the anonymous referee for a constructive report that has significantly improved this paper. We acknowledge the financial support from the National Natural Science Foundation of China under grants No. 11503078, 11573060 and 11661161010.

\bibliography{shadow_BH_DM_v2}

\end{document}